\documentstyle[aps,pre,floats,epsfig]{revtex}

\def\be{\begin{equation}}
\def\ee{\end{equation}}
\def\bea{\begin{eqnarray}}
\def\eea{\end{eqnarray}}

\begin{document}

\title{Inertia, coarsening and fluid motion in binary mixtures}

\author{M. E. Cates$^a$, V. M. Kendon$^a$, P. Bladon$^a$, J.-C. Desplat$^b$}

\address{$^a$Department of Physics and Astronomy and $^b$Edinburgh Parallel
Computing Centre, \\
University of Edinburgh, King's Buildings, Mayfield Road,
Edinburgh EH9 3JZ, Scotland}

\maketitle

\begin{abstract}\noindent Symmetric binary fluids, quenched into a regime
of immiscibility,
undergo phase separation by spinodal decomposition. In the late stages, the
fluids are separated by
sharply defined, but curved, interfaces: the resulting Laplace pressure
drives fluid flow. Scaling
ideas (of Siggia and of Furukawa) predict that, ultimately, this flow
should become turbulent as
inertial effects dominate over viscous ones. The physics here is complex:
mesoscale simulation
methods (such as Lattice Boltzmann and Dissipative Particle Dynamics) can
play an essential role in
its elucidation, as we describe. Likewise, it is a matter of
experience that immiscible fluids will mix, on some lengthscale at least,
if stirred vigorously
enough. A scaling theory (of Doi and Ohta) predicts the dependence of a
steady state domain size on
shear rate, but assumes low Reynolds number (inertia is neglected). Our
preliminary simulation results (three-dimensional,
so far only on small systems) show little sign of the kind of
steady state envisaged by Doi
and Ohta; they raise instead the possibility of an oriented domain texture
which can continue to
coarsen until either inertial effects, or (in our simulations) finite size
effects, come into play.
\end{abstract}

\section{Introduction}
When an incompressible binary fluid mixture is quenched far below
its spinodal temperature, it will phase separate into domains of
the two fluids. The simplest case (in theory at least) is when the two
fluids are thermodynamically
and kinetically symmetrical: fluids A and B have identical properties in
all respects except
that they are mutually phobic. For a deep quench (to a temperature well
below the critical point)
the thermodynamic equilibrium state then has the two fluids completely demixed.
One remaining control parameter is the volume fraction $\phi$; only for a
50:50 mix ($\phi =
0.5$, which with thermodynamic symmetry is a quench through the critical
point) is the system
totally symmetrical.  Other significant parameters are the surface tension
$\sigma$, viscosity
$\eta$ and the fluid mass density $\rho$. There is also a mobility
parameter $M$ which controls the
collective diffusion of concentration fluctuations: this can remain
important at late
times, but only if $\phi$ is small enough (less than about 0.15--0.20 for
3-D) that
the fluid domains
depercolate. The minority droplets can then continue to coarsen by a
ripening mechanism (controlled
by
$M$) or by diffusion and coalescence of the droplets themselves; both give
a mean droplet
size that scales with time as $t^{1/3}$.
If instead the domains remain connected, then the late-stage growth
is driven by capillary forces (Laplace pressure), arising from curvature of
the (sharply defined)
interface between the two fluids; these drive fluid flows from regions of
tight curvature (necks,
narrow liquid bridges) into those of low interfacial curvature (large domains).
Note that for coarsening of a bicontinuous structure to proceed this way,
one also requires
discrete ``pinch-off" events to continually occur: each of these allows
the topology to change
discontinuously in time. For a theory-oriented review of the late stages of
spinodal domain growth,
see Bray
\cite{bray}; there are many relevant experimental studies as well
\cite{experiments}.

In what follows, we address two issues concerning domain growth in three
dimensional binary fluid systems. (The two dimensional case has some
special features of its own;
see
\cite{wagner}.)  The first issue concerns the role of inertia in the late
stages; this is a
matter of continuing interest and controversy \cite{grant}, that our recent
simulation work
\cite{ourprl} illuminates. The second concerns the
effect of an applied shear flow on the coarsening process. Here we have
only preliminary
simulation results, but these are enough to suggest that there is more to
this problem than a recent
scaling theory
\cite{doi} suggests; again, our discussion focuses on inertial effects.

Our simulation work uses two different mesoscale simulation methods. One is
the Lattice Boltzmann
(LB) method, in which a velocity distribution function $f({\bf v})$ and
composition variable $\phi$
are defined at discrete lattice points. The rules for updating these
quantities represent a
discretization of the (zero temperature) Navier-Stokes equation for fluid
flow, coupled to a
Cahn-Hilliard type thermodynamic free energy functional ${\cal F}[\phi]$,
which provides the
driving force for phase separation. The second method is DPD (dissipative
particle dynamics), which
is a long-time-step, noisy molecular dynamics algorithm (off lattice)
involving soft repulsive
interaction potentials between two types of particle. (The repulsive
interactions are chosen to
favour demixing.)  This is a fixed-temperature algorithm, in which
stability is ensured, despite
the long time-step, by introducing local damping and noise terms in accord
with a suitable
fluctuation dissipation theorem. Crucially, both damping and noise act
pairwise on
particle velocities and hence conserve momentum locally, recovering the
isothermal Navier
Stokes equation at large length scales. For further information on LB for
binary fluids see
Swift et al. \cite{swift} and on DPD see Groot and Warren
\cite{groot}; details of our algorithms and parameter settings are recorded
elsewhere
\cite{ourprl,ludwig,jury}.  Note that in both methods we have run tests to
check that, under the
particular simulation conditions we adopt, fluids are behaving incompressibly.

\section{Binary demixing at high Reynolds number}

For simplicity, we restrict attention to fully symmetric mixtures ($\phi =
0.5$) for which the
fluid domains comprise, in three dimensions, a fully bicontinuous
structure. The late-time
evolution of this structure remains incompletely understood despite
theoretical \cite{siggia,furukawa,bray,grant}, experimental
\cite{experiments} and simulation
\cite{laradji,lebowitz,jury,footlin} work over recent years.

\subsection{Scaling expectations}
As emphasized by Siggia \cite{siggia}, the coarsening of a bicontinuous
spinodal texture
involves capillary forces (governed by the interfacial tension, $\sigma$),
viscous dissipation
(governed by the fluid viscosity $\eta$), and fluid inertia (governed by
the mass density $\rho$).
Out of these three physical parameters, ($\sigma$, $\eta$, and $\rho$),
only one length, $L_0 = \eta^2/\rho\sigma$ and one time
$T_0 = \eta^3/\rho\sigma^2$ can be constructed, which allow us to describe the
time evolution of the coarsening system in unique dimensionless length and
time measures.
For convenience we define the lengthscale
$L(T)$ of
the domain structure at time $T$ via the structure factor $S(k)$ as
$L = (1/2\pi)  \int S(k) dk /\int k S(k) dk$.
Provided no other physics except that described by the three 
macroscopic parameters, $\sigma$, $\eta$, and $\rho$, is
involved in late stage growth this leads us to the
dynamical scaling hypothesis
\cite{siggia,furukawa}:
\begin{equation}
l = l(t)
\end{equation}
where we use reduced time and length variables, $l \equiv L/L_0$
and $t \equiv (T-T_{int})/T_0$.  Since dynamical scaling should hold only after
interfaces have become sharp, and transport by molecular diffusion
ignorable, we have allowed for a nonuniversal offset $T_{int}$;
thereafter the scaling function $l(t)$ should, in principle,
approach a universal form, the same for all (fully symmetric,
deep-quenched, incompressible) binary fluid mixtures.

It was argued further by Furukawa \cite{furukawa} that, for small enough
$t$, fluid inertia is negligible compared to viscosity, whereas for large
enough $t$ the reverse is true. Dimensional
analysis then requires the following asymptotes:
\begin{eqnarray}
l \to b t \mbox{\hspace{1.25em}} & \mbox{\hspace{1em};\hspace{1em}} & t\ll
t^* \label{viscous}\\
l \to c t^{2/3} & \mbox{;} & t\gg t^* \label{inertial}
\end{eqnarray}
where, if dynamical scaling holds, amplitudes $b,c$ and the crossover time
$t^*$ (defined, for
example, by the intersection of asymptotes on a log-log plot) are universal.

Several earlier numerical studies claim to see one or other of these
scaling regimes, but of these,
few provide accurate values of $\eta, \rho, \sigma$ as required if the
various datasets are to be
compared on a scaling plot \cite{footlin}. Those that do so, and which
appear to confirm
Eq.\ref{viscous}, include the work of Laradji et al. \cite{laradji}, for
which $2 \le l \le 20$,
that of Bastea and Lebowitz
\cite{lebowitz} ($1 \le l \le 2$) and that of Jury, Bladon, Krishna and Cates
\cite{jury} ($20 \le l \le 2000$). Jury et al. performed a careful
comparison of their
own (DPD) and these others' datasets; while linear scaling was reported,
they found no consistency
in the value of $b$. Instead, a systematic trend appeared, in which $b$
drifts downward between
datasets as one moves to larger
$t$ and $l$ (roughly, $b \sim t^{-0.2}$). Despite this, within each DPD
dataset, the linear $t$
dependence is clearly better than a fit to  $l \sim t^{0.8}$.  Jury et al.
proposed that a
non-scaling behaviour of this kind could perhaps be explained if some
nonuniversal physics (that
is, not contained in $\sigma,\rho,\eta$) were to intervene: they suggested
a candidate involving
the physics of topological reconnection, a process that, even at late
times, could involve
a molecular (or discretization) scale, small compared to $L$.

\begin{figure}[htb!]
\begin{center}
\leavevmode
\begin{minipage}{0.47\textwidth}
        \resizebox{\textwidth}{!}{\rotatebox{-90}{\includegraphics{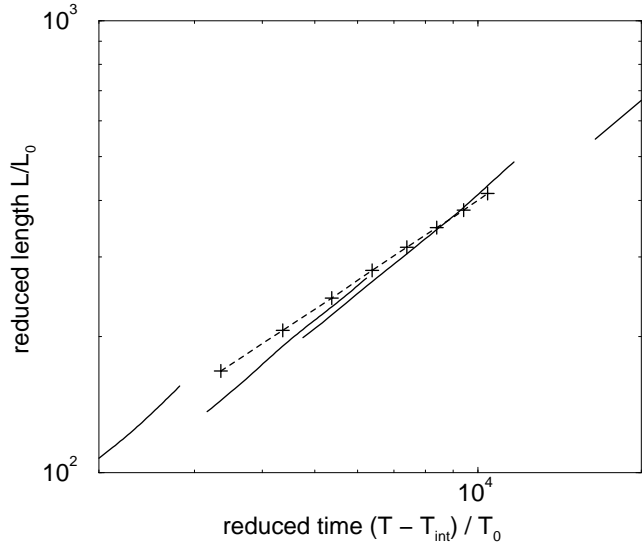}}}
	\caption{Comparison of individual $l(t)$ datasets, on a log-log
plot, generated by the DPD
\protect\cite{jury} and the LB \protect\cite{ourprl} algorithms.
LB dataset (crosses) has $L_0=0.15$ lattice units, DPD datasets (solid)
have $L_0=0.29$, 0.19, 0.13, 0.077 DPD units
(left to right).}
        \vspace{0.2cm}
        \protect\label{graph:compare}
\end{minipage}
\end{center}
\end{figure}

This possibility remains open,
although our own, more recent, LB data \cite{ourprl} suggests the following
alternative
explanation: (a) the linear law is indeed obeyed for $l \le 20$ (with $b
\simeq 0.07$), but the
$b$ coefficients of Laradji et al. and of Bastea and Lebowitz are both
overestimated due to residual
diffusion effects; (b) the data of Jury et al. lies within a broad
crossover region between
Eq.\ref{viscous} and
Eq.\ref{inertial}, where the local slope on a log-log plot is around $0.8$;
(c) the preference for
linear fits in these DPD datasets is partly caused by finite size
corrections within each
dataset. To better eliminate the latter, in our LB coarsening data we
insist that $L \le\Lambda/4$,
with $\Lambda$ the linear system size (Jury et al. allowed $L \le
\Lambda/2$). A comparison between
representative LB and DPD datasets, in a regime where both are available,
is shown in
Fig.~\ref{graph:compare}. Though not identical, it is hard to be sure that the
remaining discrepancies do not arise from finite size corrections.

\subsection{Inertial effects}

The relative importance of inertial to viscous terms, in fluid mechanical
problems, is
traditionally measured by a Reynolds number. In the spinodal context this
is usually defined as
Re = Re$_s \equiv (\rho L/\eta) dL/dT = l\dot l$. In the linear regime it
is of order $b^2t$. Note
that since
$b$ is rather small, the maximum Re achieved by Jury et al (for $t \simeq
2000$) is only about 20
even with $t$ is of order $10^4-10^5$. Indeed, this is roughly where we now
believe $t^*$ to lie;
a Reynolds number of 20 is large enough for inertial effects to
be non-negligible, but by fluid mechanics standards, still modest. Note
also that the
scaling ideas clearly predict that Re should increase without
bound (as $t^{1/3}$) within the inertial regime, Eq.\ref{inertial}.
However, in a recent paper \cite{grant} Grant and Elder have argued
that the Reynolds number cannot, in fact, continue to grow
indefinitely. If so, Eq.\ref{inertial} is not truly the large $t$
asymptote,
which must instead have $l\sim t^\alpha$ with $\alpha \le \frac{1}{2}$.
Essentially,
Grant and Elder argue that at large enough Re, turbulent remixing of the
interface
will limit the coarsening rate \cite{grant}, so that Re stays bounded at a
level
which they estimate as Re $\sim 10-100$.

\begin{figure}[htb!]
\begin{center}
\leavevmode
\begin{minipage}{0.47\textwidth}
        \resizebox{\textwidth}{!}{\rotatebox{-90}{\includegraphics{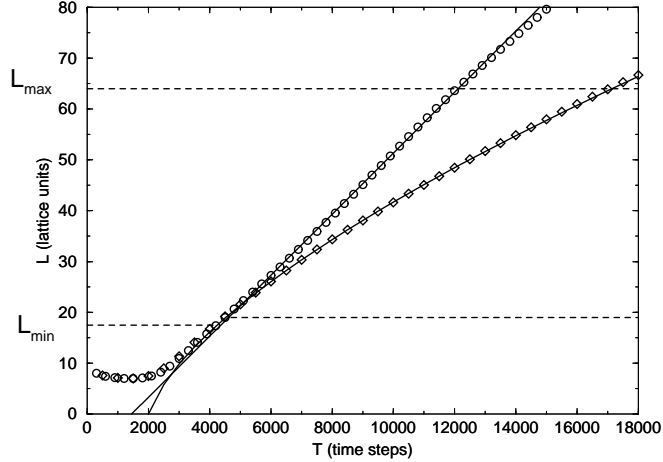}}}
	\caption{$L$ vs. $T$ (in lattice units) for runs with
		 $L_0$ = 5.9 (circles) and 0.0003 (diamonds).
		 Dotted lines delineate the range of data points used
		 for fitting ($17<L<64$) and the fits to $t^1$ and $t^{2/3}$
		 respectively
		 have been projected back to show the intercepts,
		 $T_{int}$.}
        \vspace{0.2cm}
        \protect\label{graph:raw-data}
\end{minipage}
\end{center}
\end{figure}

Our recent LB work \cite{ourprl} represents the first large-scale
simulations of 3-D spinodal
decomposition to unambiguously attain a regime in which inertial forces
dominate
over viscous ones, allowing a test of this idea. We find direct evidence
for Furukawa's
$l\sim t^{2/3}$ scaling, Eq.\ref{inertial}. Although a further crossover to
a regime
of saturating Re cannot be ruled out, we find no evidence for this up to Re
$\simeq
350$.
Two of our LB datasets are shown in
Fig.\ref{graph:raw-data}. The first has high $L_0$ (high viscosity) and
corresponds to
$l$ values around unity ($t\simeq 10$) for the fitted part of the data. It
is well fit by
Eq.\ref{viscous}. The second has low $L_0$ and corresponds to $l$ around
$10^5$
(or $t\simeq 10^7$); it is well fit by Eq.\ref{inertial}. The Reynolds
number is about 0.1
near the end of the first run and about 350 near the end of the second.

\begin{figure}[htb!]
\begin{center}
\leavevmode
\begin{minipage}{0.47\textwidth}
        \resizebox{\textwidth}{!}{\rotatebox{-90}{\includegraphics{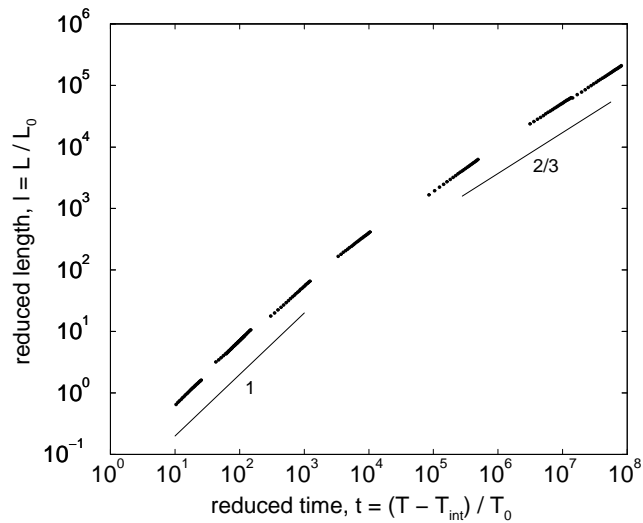}}}
        \caption{Scaling plot in reduced variables $(l,t)$ for eight LB
                 datasets.  Second and third from left overlap and 
                 cannot be distinguished on the plot.}
        \vspace{0.2cm}
        \protect\label{graph:cooked}
\end{minipage}
\end{center}
\end{figure}

It goes without saying that
no {\em single} simulation run could possibly cover this range of length or
time scales, since that
would require a lattice of around $\left(10^6\right)^3$ sites. Our lattices
are $256^3$, but by
changing the LB parameter values, we can access both regimes, and
everything in between
\cite{ourprl}. Similar exploration is possible (to a lesser extent) within
DPD \cite{jury}. Such
mesoscale methods have strong advantages over (say) MD
\cite{laradji} since, after proper calibration and subject to specified
range limitations, they
allow one to ``dial in'' ones own choice of thermodynamic ($\sigma$) and
kinetic ($\eta$)
parameters. Thus one can build up the $l(t)$ curve section by section; if
it is universal, this is
enough.  (For a fuller explanation of the procedure, see Kendon et al.
\cite{ourprl}.)
This is done for our LB data
in Fig.\ref{graph:cooked}; note that the rightmost two datasets (both in
the inertial regime) are
almost contiguous, as one would hope on a universal scaling plot. (Each has
a slope close to $2/3$.)
However, computational restraints prevent us so far from covering the
entire $l(t)$ curve with data;
it remains possible that some mismatches between runs, of the type reported
by Jury et al.
\cite{jury} might still be observed at intermediate $l,t$. Note the broad
crossover between viscous
(three leftmost runs) and inertial (two rightmost runs) regimes: this
crossover covers four decades
in $t$ (or three in $l$). But it is less spectacular when expressed in
terms of Re; our data span
$0.1 \le$ Re $\le 350$ and the crossover region is roughly Re $= 1-100$.

\begin{figure}[htb!]
\begin{center}
\leavevmode
\begin{minipage}{\textwidth}
       \resizebox{\textwidth}{!}{\includegraphics{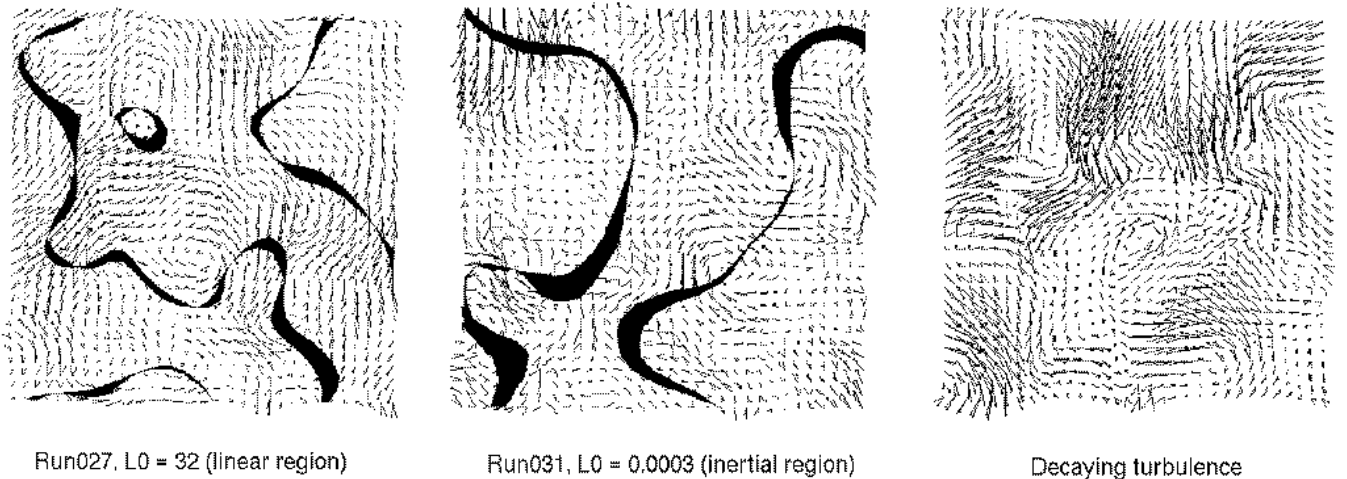}}
        \caption{Velocity maps for $L_0 = 5.9, 3.0 \times 10^{-4}$ (LB),
	         and for freely decaying isotropic turbulence
	         (pseudo-spectral dns).
		 Each map projects a $32\times 32\times 2$
		 thin section taken from a $128^3$ grid.
		 The interface appears wide where it cuts the section
		 at a glancing angle; each site shows two arrows which
		 are the projected velocity vectors
	 	 in the two layers of the section.}
        \vspace{0.2cm}
        \protect\label{graph:velmaps}
\end{minipage}
\end{center}
\end{figure}

It is important now to ask whether, at the largest Re values we can reach,
there is significant
turbulence in the fluid flow.  One quantitative signature of turbulence is the
skewness
$S$ of the longitudinal velocity derivatives; this is close to zero in
laminar flow but approaches $S=-0.5$ in fully developed turbulence
\cite{turbulence}.
A space-averaged value close to $S=-0.5$ would imply turbulence
everywhere and, presumably, significant remixing at the interface
\cite{grant}. We do detect increasingly negative $S$ as Re is increased but
reach only
$S\simeq -0.35$ for Re $\simeq 350$ \cite{long}. This suggests that at
our highest Re's,  partially but not fully developed turbulence is present.
This view is consistent
with Fig.\ref{graph:velmaps}, which shows velocity maps for low and high Re
runs, and (for
comparison) single fluid freely decaying isotropic turbulence.
In a low Re flow one expects the only relevant lengthscale in the system
to be the domain size $L$; at high Re, there should be a cascade of
structure in the velocity field
below the domain size. Some internal structure is plausibly, though not
conclusively, visible in the
plot for our lowest $L_0$ value. But since we do not yet see fully
developed turbulence,
it remains an open issue whether such effects would
lead, beyond the observed $t^{2/3}$ inertial scaling regime, to a final,
turbulent remixing regime
(of saturating Re) as proposed by Grant and Elder
\cite{grant}. If it does, the limiting value of Re must significantly exceed
their estimate of
$10-100$.

\section{Demixing under flow}
Consider now the situation where a symmetric binary fluid undergoes a deep
quench in conditions
where a steady shear flow is applied externally. In simulations, there are
two ways to achieve
this. One is by Lees-Edwards (moving periodic) boundary conditions, which
have been implemented in
our DPD codes \cite{pccp}; the other is to have sliding solid boundaries,
and the LB results
reported below use this method.
For binary fluids under shear, the two approaches give
qualitatively similar results,  For related work in two and three
dimensions, see \cite{shear_2D,AW}.
The questions are as follows: does the
presence of flow arrest the
coarsening process, giving a steady state with a finite domain size,
and, if so, what is the character of this steady state?  Also, does this
steady state depend on
starting conditions -- for example, if the binary fluids are allowed to
demix first and the flow is
then switched on, is the same final state achieved?  Perhaps surprisingly
(despite the obvious
importance of mixing and emulsification in many technological areas)
well-founded theoretical
answers to these questions are almost nonexistent.

\subsection{The Doi-Ohta theory}
An important theoretical inroad into this problem was made by Doi and Ohta
\cite{doi}. (Its
rheological consequences are reviewed in the book by Larson \cite{larson},
where comparisons to
experiment can be found.)  The Doi-Ohta work is entirely limited, however,
to a regime of
vanishingly small Reynolds number. Note that for this problem, the relevant
Re is the larger of
Re$_s= (\rho L/\eta) dL/dT$ (the value extracted from the coarsening
process, which must vanish in
any true steady state) and
\begin{equation}
\hbox{\rm Re}_f = \rho\dot\gamma L^2/\eta \label{flowren}
\end{equation}
which is the value found by considering the effect of the external shear
rate $\dot\gamma$ on a
droplet of scale
$L$. Both definitions assume, perhaps implausibly, that there is only one
relevant length, $L$, which
requires, for example, that domains are not strongly anisotropic; we return
to this point below.

Doi and Ohta assumed that, for nonzero $\dot\gamma$, a steady state would
be reached at negligible
Re (so that inertial effects can be ignored). The applied flow stretches
and deforms the
interfaces (tending also to orient these along the flow direction); this is
opposed by interfacial
tension. The competition is measured by the capillary number Ca $= T_0\dot
\gamma L/L_0
= \dot \gamma L \eta/\sigma$; at small Ca, a spherical droplet of size $L$
is weakly perturbed,
but at large Ca it should be elongated to the point of rupture.
By estimating the magnitude of each effect in the absence of the other, and
balancing
these, Doi and Ohta made the intriguing prediction that, in steady state,
the fluid should have a
$\dot\gamma$-independent viscosity of the form $\eta f(\phi)$. That is, the
behaviour should be
independent of $\sigma$ and $\rho$, but dependent on $\phi$
(values of $\phi$ close to 0.5 were assumed.) This form follows purely by
dimensional analysis if
one insists that the only relevant variables are $\dot\gamma, \sigma,
\eta$ and $\phi$ (the absence of $\rho$
from this list is because inertia is neglected.) Doi and Ohta went on to
develop specific
approximation schemes that allowed them, for example, to
study transient (shear startup) behaviour.

In the steady state, the departure of $f(\phi)$ from unity must come from
the presence of
interfacial stresses, which, if the structure is characterized by a single
domain length $L$, scale
as $\sigma/L$. Comparing this with the excess shear stress $\eta \dot\gamma
(f-1)$ we obtain a
length estimate in steady state
\begin{equation}
L \simeq {\sigma\over\eta\dot \gamma} \label{length}
\end{equation}
which corresponds to the system  arranging itself, in steady state, so that
the capillary number is
of order unity.
This argument is, of course, somewhat oversimplified. If more than one length
scale is present (for example, describing the structure along different
directions relative to the
flow) then one has to determine which one dominates the interfacial
contribution to the viscous
stress. The largest Laplace stresses arise in regions of strongest
curvature;  however, if the
strongest curvature is axial about the flow direction (for example, thin
cylinders lying along the
flow) it cannot contribute to the shear stress at all \cite{doi}.
Doi and Ohta did not
much explore these geometrical aspects of their work. Note that, implicit
in their
whole description, is an assumption that a steady state can be achieved
without inertial terms
coming into play. Our preliminary results on sheared spinodal systems,
given below, cast some doubt
on this assumption.

\subsection{Results}
We have simulated sheared binary fluids using both DPD and LB algorithms.
The results presented
here are for relatively small systems (50,000 DPD particles, and $64^3$ or
$96^3$ lattices) and thus
are preliminary, but we think they point towards some interesting physics.
In future work we hope to
address larger systems; among other things, this will allow reliable
statistics for the interfacial
stresses, and hence for the rheological properties of the sheared systems,
to be obtained.

\begin{figure}[htb!]
\begin{center}
\leavevmode
\begin{minipage}{0.47\textwidth}
        \resizebox{\textwidth}{!}{\rotatebox{-90}{\includegraphics{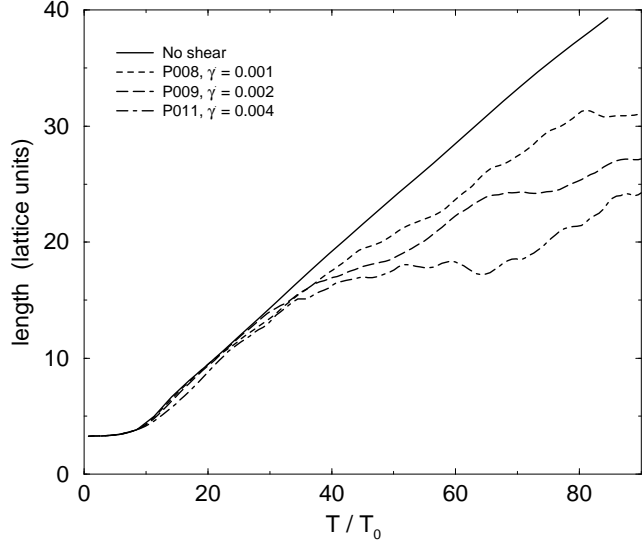}}}
  \caption{
Overall coarsening of the domains as estimated from the amount of
interface in an LB system for shear rates of $\dot{\gamma} T_0$ = 0.07,
0.14, 0.21, compared to zero shear.  Using the system size to estimate the
maximum capillary number for the systems due to the shear
gives Ca = 0.4, 0.8, 1.5.}
        \vspace{0.2cm}
        \protect\label{graph:onelength}
\end{minipage}
\end{center}
\end{figure}

Our first result is that shear does inhibit coarsening. If one defines a
single lengthscale via the
total surface area $A$ in the system (so that $A = \Lambda^3/L$),
then indeed this lengthscale increases more slowly, when shear is applied,
than without it
(Fig.\ref{graph:onelength}).

\begin{figure}[htb!]
\begin{center}
\leavevmode
\begin{minipage}{0.47\textwidth}
        \resizebox{\textwidth}{!}{\rotatebox{-90}{\includegraphics{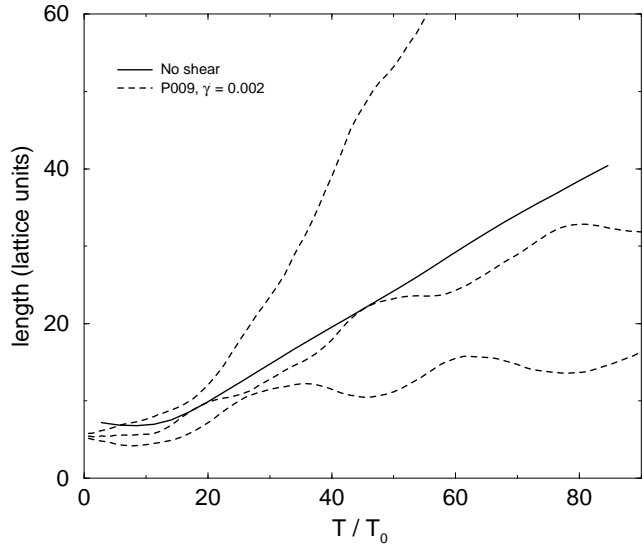}}}
 \caption{
Time evolution of three length scales given by the eigenvalues of the
curvature matrix
for the LB system with $\dot{\gamma} T_0$ = 0.14. }
        \vspace{0.2cm}
        \protect\label{graph:threelengths}
\end{minipage}
\end{center}
\end{figure}

On the other hand, this one-lengthscale analysis fails to reflect the
extreme anisotropy that can
develop. Indeed,  by examining the eigenvalues $\lambda_1$, $\lambda_2$,
$\lambda_3$,
of the interfacial curvature matrix\cite{AW},
\begin{equation}
D_{\alpha\beta} = \frac{\sum\limits_{grid}
\partial_{\alpha}\hat{\phi}\partial_{\beta}\hat{\phi}}
		       {\sum\limits_{grid} \hat{\phi}^2},
\end{equation}
lengths for the three different directions can be estimated,
\begin{equation}
L_1 = \frac{\xi}{\lambda_1}, \mbox{ }
L_2 = \frac{\xi}{\lambda_2}, \mbox{ }
L_3 = \frac{\xi}{\lambda_3}.
\end{equation}
In these expressions, $\xi$ is the equilibrium width of the fluid-fluid
interface, and $\hat{\phi}$
is the local compositional order parameter defined on the lattice. The time
evolutions of these
three lengths  are plotted for a system with $\dot\gamma T_0 = 0.14$ in
Fig.\ref{graph:threelengths}.

\begin{figure}[htb!]
\begin{center}
\leavevmode
\begin{minipage}{\textwidth}
       \resizebox{0.31\textwidth}{!}{\includegraphics{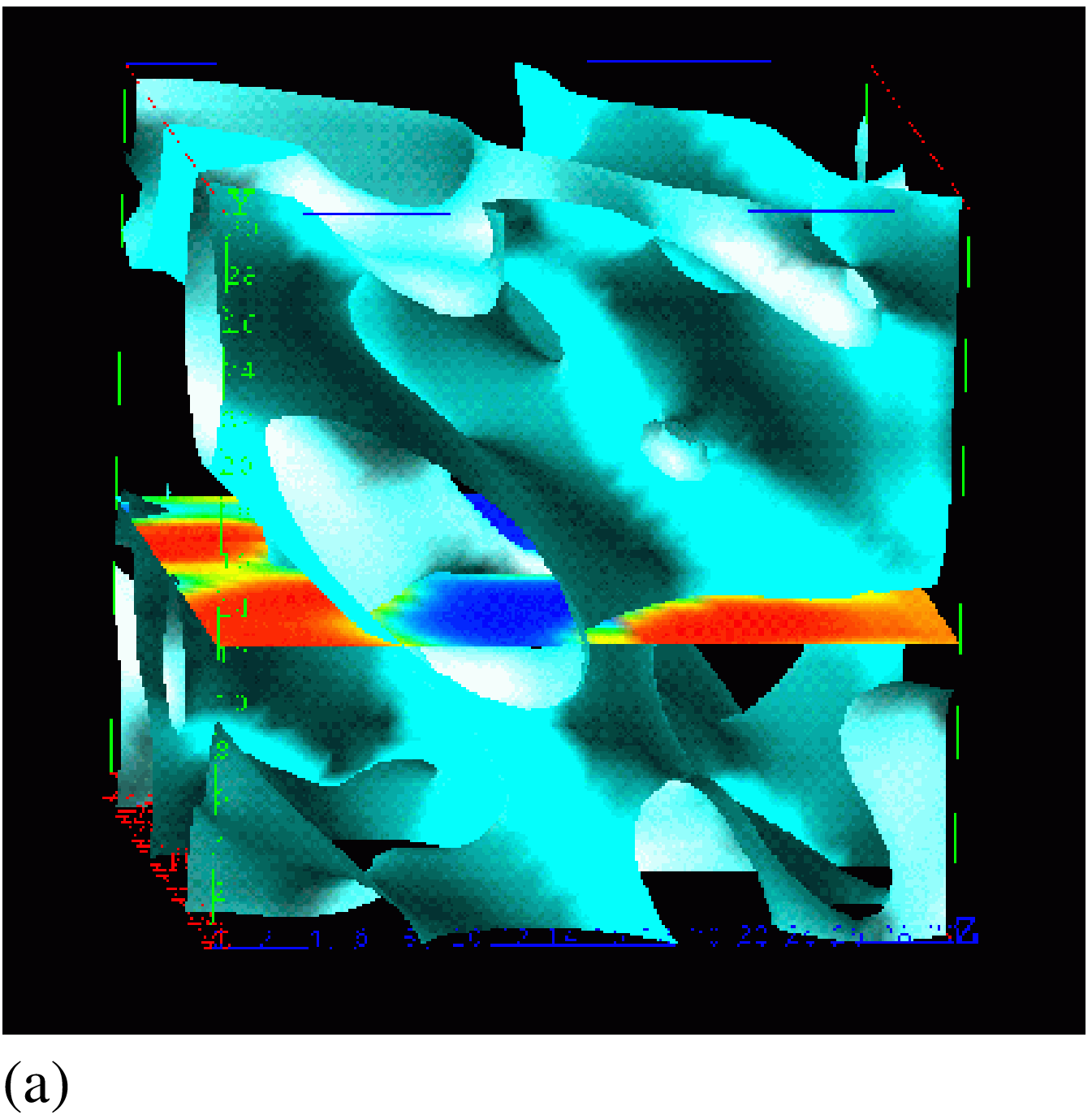}}
	\rule[0cm]{0cm}{5.5cm} \hfill
       \resizebox{0.302\textwidth}{!}{\includegraphics{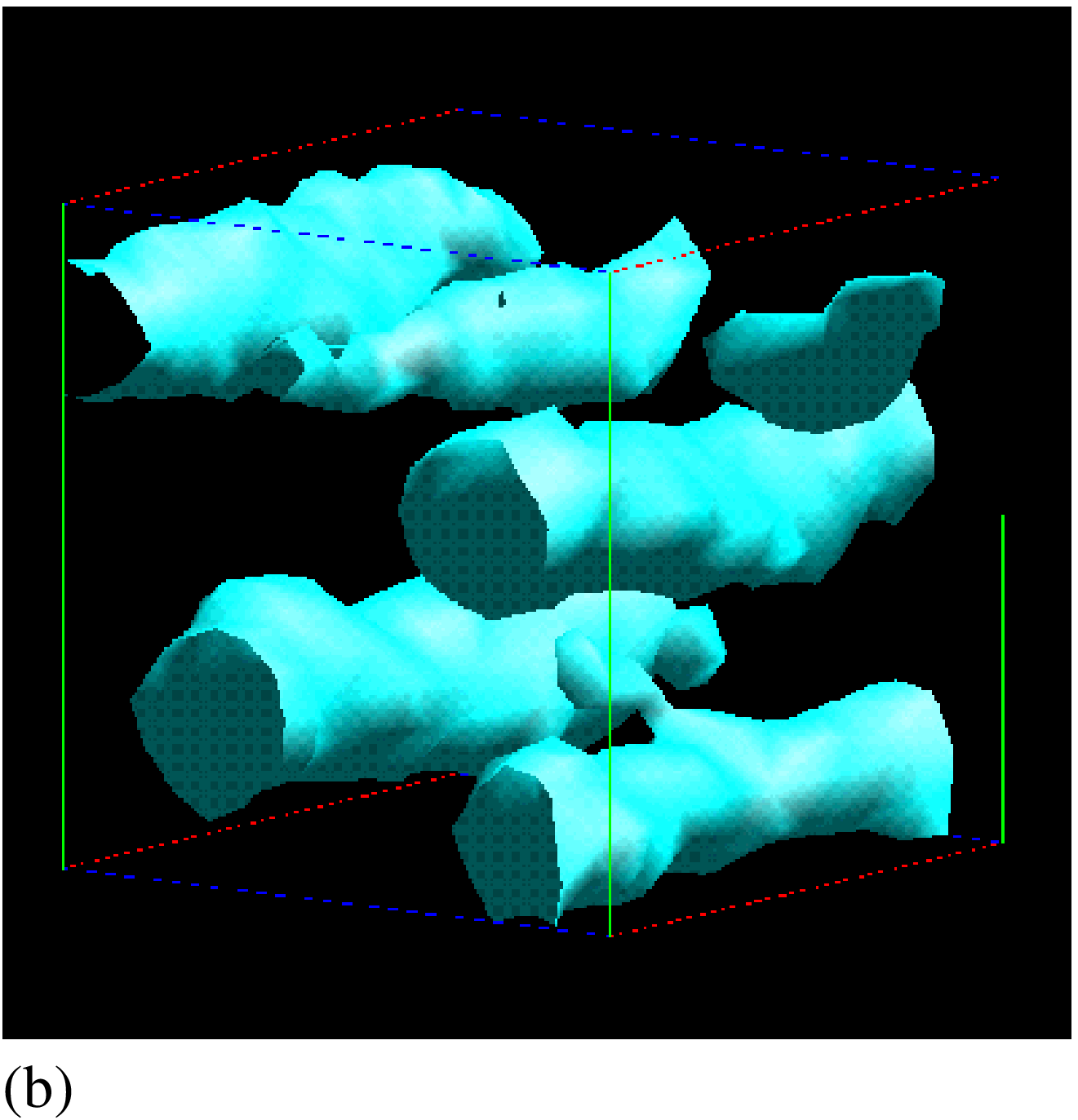}}
	\hfill
       \resizebox{0.32\textwidth}{!}{\includegraphics{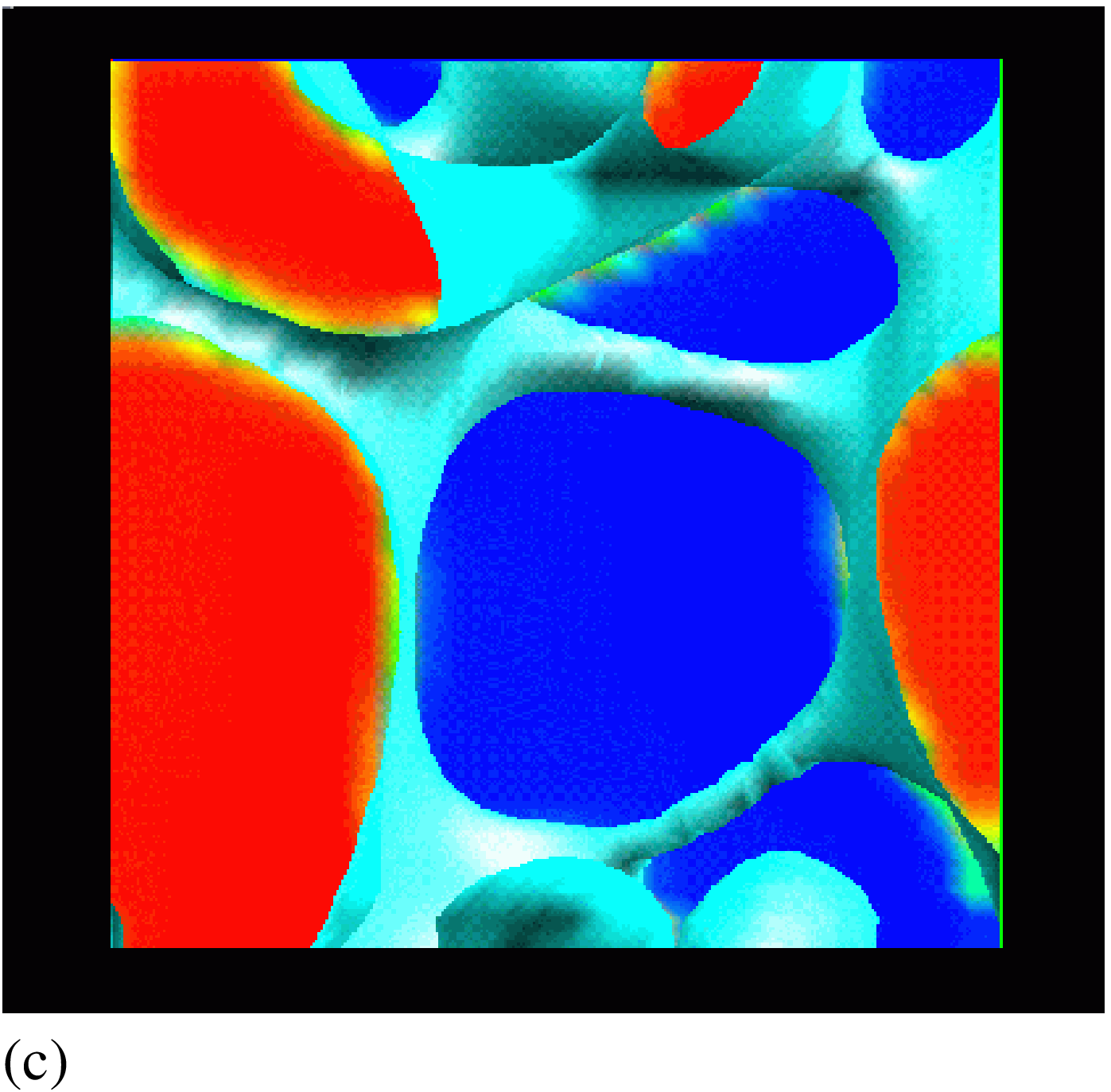}}
\vspace{0.2cm}
 \caption{(a) LB simulation with $\phi = 0.5$, prior to steady state being
achieved.  Shown is a central $32^3$ section from a $64^3$ system,
ith the shear applied through the top plane moving left
and the bottom plane moving right.
Light blue denotes the fluid-fluid interface.
The horizontal plane shows a slice through the order parameter profile,
with solid colours representing the two
fluids: this shows the interfacial sharpness. (b) DPD simulation with
$\phi = 0.2$ and
$\dot\gamma T_0 = 0.32$. Tubes are oriented along the streamlines. (c) LB
simulation with $\phi = 0.5$ and
$\dot\gamma T_0 = 0.14$.  This view is along the streamlines; the top and bottom
edges are the moving
boundaries. The  block colours (as in (a)) show the identity of the two
fluids, now on a
vertical plane towards the back of the simulation cell. Passing in front of
this, one observes
narrow fluid necks connecting blocks of similar fluid.}
        \vspace{0.2cm}
        \protect\label{graph:tubes}
\end{minipage}
\end{center}
\end{figure}

This data shows strong anisotropy developing, with elongation of domains
along the flow direction.
Under these conditions, finite size effects become greatly emphasized
\cite{wiggle}:
domains, elongated along
the flow, soon reach the size of the simulation box and connect up to their
periodic images.
A sheared LB configuration, during the initial phase before this happens
(such that $L_{1,2,3}$
are all small compared to the simulation size $\Lambda$) is shown in
Fig.\ref{graph:tubes}(a).
However, this configuration is not close to a steady state.

The structure we observe on approach to the steady state is, in all cases
we have so far studied,
clearly finite-size influenced. For
$\phi$ different from 0.5, the appearance is like that of a ``string phase"
\cite{string} in
colloids: effectively infinite tubes of the minority fluid oriented along
the streamlines. A
typical example of this, found using DPD, is shown in
Fig.\ref{graph:tubes}(b). Although no
longer bicontinuous in the transverse direction, this structure can
continue to coarsen, very
slowly, by diffusive transport (ripening): it seems likely that, for this
system size, the final
state will be a single cylinder of fluid. Figure \ref{graph:tubes}(c) shows
a LB
simulation for $\phi = 0.5$ viewed along the streamlines; this has
coarsened to the point where
finite size effects are obvious even transverse to the flow direction. The
structure is strongly
oriented, with continuous domains of the two fluids passing right through
the sample from front to
back; however, it remains connected transverse to the flow with thin
necklike bridges clearly
visible. In fact, for $\phi = 0.5$, such transverse connectivity is
inevitable unless the symmetry
between the two fluids is spontaneously broken. Thus the structure remains
capable of
coarsening further, by a non-diffusive hydrodynamic mechanism. For this system size, the
final state could be
total separation, just as it would be without shear flow: such a state
might be stable so long as
the final interface between the two fluids is parallel to the streamlines.

\subsection{Discussion}
In our simulations on these (fairly small) systems, we see no evidence of a
steady state structure
emerging that is independent of system size. This is intriguing. If the
Doi-Ohta scaling is
correct, we should obtain a length scale given by Eq.\ref{length}; so long
as this is small
compared to $\Lambda$ (and the steady state structure is not too
anisotropic) the simulation should
achieve a proper steady state, representative of that of an infinite
sample. Put differently, this
will arise if the maximum capillary number attainable in a system of size
$\Lambda$, Ca
$=\dot\gamma \Lambda \eta/\sigma$, is much larger than one.
This is true for our largest DPD runs, with $\mathrm{Ca} \sim 7$,
yet we see no sign of any 
steady state not dominated by finite size effects; larger runs would
be useful to clarify this issue.

An important possibility, which these results suggest, is as follows. By
neglecting inertia and yet assuming a steady state, Doi and Ohta implicitly
assume that this
steady-state can be characterized by a notional capillary number Ca
$=\dot\gamma L\eta/\sigma$
(based on a single measure of domain size $L$) of order one. This might be
incorrect: without
inertia, there may simply be no obstacle to complete separation of the two
fluids in simple shear.
Indeed, one can instead postulate a fully phase separated state with two
blocks of fluid, separated
by an interface in the plane of shear (or, in fact, any other orientation
with the surface normals
everywhere perpendicular to the fluid velocity). For such a flow, the {\em
notional} capillary
number is infinite (since $L$ is infinite), but the {\em effective} value
is zero: the
interface is oriented parallel to the streamlines and there is no
stretching of it by the flow.
This structure is likely to be destabilized by inertial contributions,
since it involves steady
shear of a laminar interface between two fluids (although this may require
a threshold of shear rate, such as $\dot\gamma T_0 \simeq 1$, to be exceeded).
Thus it remains possible that a steady state, with a non-infinite domain
scale can arise {\em only}
when inertial terms are significant. If so the Reynolds number is just as
important as the capillary
number, and the implicit criterion Ca $\simeq 1$ of the Doi-Ohta theory is
not adequate.
To test this point,  one needs to run simulations under shear at high Re. We
intend to pursue this, using our parallel LB code \cite{ludwig}, in the
near future.

Meanwhile it is straightforward to check that, at the system sizes and
shear rates used here, a
starting configuration with oriented slabs of fluid (with an interface
parallel to the shear planes)
remains essentially unperturbed by the flow. However, if the initial
interface normal is not
perpendicular to the streamlines, the slabs are immediately broken up, and
the late stage
coarsening (with a slowly-evolving, tubular structure along the
streamlines) resembles that which we
found above for the case of an initially homogenous phase, quenched to form
demixing domains,
in the presence of steady shear.

\section{Conclusions}

We have described two problems, in the physics of immiscible binary fluids,
where inertial effects
are important. One is the late stages of demixing (in the absence of
applied flow) where the
internal dynamics of the system drives it to high Reynolds numbers. Whether
the final Re is
self-limiting, as recently suggested by Grant and Elder \cite{grant}
remains to be seen; however,
we have observed \cite{ourprl} clear evidence of an inertial-dominated
regime in which the Furukawa
($l\sim t^{2/3}$) scaling is found. The second problem is in the steady
state behaviour of binary
fluids in simple shear. Our preliminary simulation results, though far from
conclusive, inspired us
to speculata that the very existence of such a steady state is itself
dependent on inertial effects.
If so, the scaling analysis that underlies Doi-Ohta theory \cite{doi}, is
in doubt. Future
simulations on much larger systems should allow us to settle this point.

We thank Simon Jury, Patrick Warren and Julia Yeomans for valuable
discussions, and Alistair Young for the turbulence dns simulation code.
Work funded in part
under the EPSRC E7 Grand Challenge.

\end{document}